\documentclass[12pt, a4paper]{article}
\usepackage[english]{babel}
\usepackage{amssymb}
\usepackage{amsthm}
\usepackage{hyperref}
\usepackage[]{graphicx}
\usepackage[]{subfigure}

\def\be{\begin{equation}}
\def\ee{\end{equation}}
\def\beq{\begin{eqnarray}}
\def\eeq{\end{eqnarray}}
\def\a{\alpha}
\def\b{\beta}
\def\d{\delta}
\def\N{{\cal N}}
\newcommand{\Tr}{{\rm Tr}}
\newcommand{\eq}[1]{(\ref{#1})}

\title{\bf Particle Production near an AdS Crunch}

\author{Lorenzo Battarra\footnote{lorenzo.battarra@apc.univ-paris7.fr} \ and Thomas Hertog\footnote{hertog@apc.univ-paris7.fr}\\
{\small {}}\\
{\small {\it  APC (CNRS, Universit\'e 
Paris-Diderot) , 10 rue Alice Domon et L\'eonie Duquet,}}\\
{\small {\it  75205 Paris Cedex 13, France}}\\
}

\begin{document}
\maketitle

\begin{abstract}

We numerically study the dual field theory evolution of five-dimensional asymptotically anti-de Sitter solutions of supergravity that develop cosmological singularities. The dual theory is an unstable deformation of the ${\cal N}=4$ gauge theory on $ \mathbb{R} \times S ^{3} $, and the big crunch singularity in the bulk occurs when a boundary scalar field runs to infinity. Consistent quantum evolution requires one imposes boundary conditions at infinity. Modeling these by a steep regularization of the scalar potential, we find that when an initially nearly homogeneous wavepacket rolls down the potential, most of the potential energy of the initial configuration is converted into gradient energy during the first oscillation of the field. This indicates there is no transition from a big crunch to a big bang in the bulk for dual boundary conditions of this kind.

\end{abstract}

\section{Introduction}

One of the main goals of quantum gravity is to advance our understanding of the nature of the big bang. A particularly important issue is whether the big bang represents the beginning of the universe, or whether semiclassical evolution essentially continues further back in time, possibly involving a transition from a big crunch to a big bang \cite{Gasperini1993,Khoury2002}.

In recent years this problem has been studied in the context of the AdS/CFT correspondence \cite{Maldacena:1997re}, which allows one to map toy model cosmologies with anti-de Sitter (AdS) boundary conditions to a dual quantum field theory living on the boundary of spacetime. Explicit examples of collapsing $AdS$-cosmologies were constructed in $\N=8$ supergravity in four \cite{Hertog:2004rz,Hertog:2005hu} and five \cite{Craps2007} dimensions, using well-defined generalizations of the usual boundary conditions on some of the negative mass squared scalars which allow smooth, asymptotically $AdS$ initial data to evolve into a spacelike singularity that reaches the spacetime boundary in finite time. The properties of these solutions and their duals were further explored in \cite{Elitzur:2005kz,Barbon2010,Bernamonti2009,Asnin2009}. However, a clear understanding of the quantum nature of cosmological singularities is still lacking.

The dual description of the $AdS$-cosmologies found in \cite{Hertog:2004rz,Hertog:2005hu,Craps2007}  involves field theories with scalar potentials which are unbounded below that drive certain boundary scalars to infinity in finite time. 
Here we concentrate on the five-dimensional solutions, where the dual field theory is a deformation of ${\cal N}=4$ Super-Yang-Mills (SYM) theory on $ \mathbb{R} \times S ^{3} $ by an unbounded double trace potential $-f {\cal O}^2/2$, with ${\cal O}$ a trace operator quadratic in the adjoint Higgs scalars. This deformation is renormalizable and the coupling $f$ that governs the instability is asymptotically free \cite{Craps2007}.

The big crunch singularity in the bulk occurs when the boundary scalar ${\cal O}$ diverges. Consistent quantum evolution requires that one imposes boundary conditions at infinite scalar field, {\it i.e.} a self-adjoint extension of the system. With a self-adjoint extension the boundary evolution is unitary and predicts a perfectly homogeneous wave packet rolls down the negative potential and bounces back. In the bulk this behavior would correspond to a quantum transition from a big crunch to a big bang, as envisioned e.g. in ekpyrotic cosmology. 

However, it has been argued \cite{Hertog:2004rz} that the full field theory evolution is likely to be very different. This is because when a homogeneous wave packet rolls down, the negative effective mass term in the potential amplifies long wavelength quantum fluctuations hereby converting the potential energy of the initial configuration into gradient energy. In order for the boundary theory to predict a transition from a collapsing phase in the bulk to an approximately homogeneous, semiclassical expanding cosmology after the singularity there must be a parameter regime where ${\cal O}$ returns close to its original value after one oscillation. However, this requires the backreaction of the inhomogeneities on the evolution of the homogeneous wave packet to be sufficiently small. 

Here we quantify the tachyonic amplification of fluctuations by numerically evolving the field theory on a lattice.
To study the question of a possible big crunch big bang transition, one is interested in the dynamics on rather short timescales and in particular in the first oscillation of the field. Since the occupation numbers of the relevant perturbation modes become rapidly very large, the system behaves classically during these first stages of the evolution. The scalar potential is unbounded below, but the self-adjoint extension essentially acts as a brick wall at infinity. We therefore model this by adding a steep regularization term to the unbounded potential and considering the limit in which the UV regulator $\gamma$ introduced this way is taken to be very small.
We find that in the $\gamma \rightarrow 0 ^{+}$  limit, the tachyonic amplification of perturbations is a very efficient mechanism to produce particles and prevents the homogeneous mode from rolling back up the potential. We conclude, therefore, that for `brick-wall' boundary conditions at infinity and for the class of models considered here, the dual description predicts no transition from a big crunch to a big bang in the bulk.

At the same time, however, our simulations provide some insight in what kind of models might exhibit a different behavior near the singularity. We discuss this briefly at the end of this paper.

\section{AdS Cosmology and its Dual Description}

\subsection{AdS Cosmology}

Our starting point is $ \mathcal{N} = 8$ gauged supergravity in five dimensions, which is thought to be a consistent truncation of ten-dimensional type IIB supergravity on $S ^5$. This admits a consistent truncation to gravity coupled to a single $SO(5)$-invariant scalar $\varphi$. The action then reduces to
\begin{equation}\label{lagr}
S = \int d ^5x \, \sqrt{-g} \left( \frac{1}{2} \mathrm{R}^{(2)} - \frac{1}{2} (\partial \varphi) ^2 + 
\frac{1}{4R _{AdS}^2} \left( 15 e^{2 \gamma \varphi} + 10 e^{-4 \gamma \varphi} - e^{-10 \gamma \varphi} \right) \right)
\end{equation}
where $ \gamma \equiv \sqrt{2/15}$ and where we have chosen units in which the five-dimensional Planck mass is unity. The maximum of the potential at $\varphi=0$ corresponds to the $AdS_5$ vacuum solution. Small fluctuations around this have $m^2=-4/ R_{AdS}^{2}$, which saturates the Breitenlohner-Freedman (BF) bound in five dimensions.

In all asymptotically AdS solutions, the scalar field $\varphi$ decays at large radial coordinate $r$ as
\begin{equation}\label{asscalar}
\varphi \sim\frac{ \alpha (t,\Omega) \ln{r}}{r ^2} + \frac{ \beta (t,\Omega)}{r ^2}
\end{equation}
where $(t,\Omega)$ are the time and angular coordinates of the boundary $\mathbb{R} \times S ^{3} $ of the $AdS$ cylinder.
To define the theory one ought to specify boundary conditions at $r=\infty$ on the metric and scalar field. This amounts to specifying a relation between $\a$ and $\b$ in (\ref{asscalar}). The usual boundary conditions correspond to taking $\a=0$, and leaving $\b$ unspecified. Alternatively, one can adopt boundary conditions of the form \cite{Hertog:2004ns}
\be \label{genbc}
\a = -\frac{\d W }{ \d \b},
\ee
where $W(\b)$ is an essentially arbitrary real, smooth function. Boundary conditions of the form (\ref{genbc}) are invariant under global time translations. The conserved energy associated with this depends on $W$, but is well-defined and finite.
In this paper we adopt scalar field boundary conditions 
of the form $\alpha = f \beta$, with $f$ an arbitrary constant. The corresponding asymptotic form of the metric and the expression of the conserved mass can be found in \cite{Hertog:2004dr,Henneaux:2004zi}. 

For small, positive values of $f$, solutions were found \cite{Hertog:2004rz} in which smooth, spherically symmetric initial data of approximately zero mass evolve into a big crunch -- a spacelike singularity that reaches the boundary of AdS in finite global time. These solutions can be viewed as a five-dimensional version of open FLRW universes in which $\varphi$ rolls down the negative potential, causing the scale factor $a(t)$ to vanish in finite time. In terms of global time $t$ it was found that $\b \sim\b(t=0)/(\cos t/R _{AdS})^2$. Hence $\b \rightarrow \infty$ as $t/R _{AdS} \rightarrow \pi/2$, which is when the singularity hits the boundary \cite{Hertog:2004rz}.

\subsection{Dual CFT description}
{\it
\subsubsection*{Classical Evolution}
}%

With the usual $\a=0$ boundary conditions, the dual field theory is $\N=4$ super Yang-Mills theory. The bulk scalars that saturate the BF bound in $AdS$ correspond in the gauge theory to the operators $\frac{1}{N} \,\Tr[\Phi^i \Phi^j - (1/6) \delta^{ij} \Phi^2]$, where $\Phi^i$ are the six scalars in $\N=4$ super Yang-Mills. The $SO(5)$-invariant bulk scalar $\varphi$ that we have kept in (\ref{lagr}) couples to the operator 
\be \label{operator}
{\cal O}=\frac{1}{N} \,\Tr\left[\Phi_1^2 - {1\over5} \sum_{i=2}^6 \Phi_{i}^2\right].
\ee

In general, imposing nontrivial boundary conditions $\a (\b)$ in the bulk corresponds to adding a multi-trace interaction $W({\cal O})$ to the CFT action \cite{Witten:2001ua,Berkooz:2002ug}, such that after formally replacing ${\cal O}$ by its expectation value $\b$ one obtains (\ref{genbc}). Hence the boundary conditions $\alpha = f \beta$ that we have adopted correspond to adding a double trace term to the field theory action
\be\label{Ocubed}
S = S_0 - W({\cal O}) = S_0+\frac{ f }{ 2} \int {\cal O}^2.
\ee
The operator ${\cal O}$ has dimension two, so the extra term is marginal and preserves conformal invariance, at least classically. 

We have taken the constant $f$ to be small and positive in the bulk. The term we have added to the CFT action, therefore, corresponds to a negative potential. Hence it is plausible that the dual field theory admits negative energy states and has a spectrum that is unbounded below. This would mean that the usual vacuum must be unstable, and that there are 
(non-gravitational) instantons which describe its  decay. After the tunneling, the field rolls down the potential and becomes infinite in finite time.  This corresponds to the big crunch singularity in the bulk and provides a qualitative dual explanation for the fact that the function $\beta$ of the asymptotic bulk solution (\ref{asscalar}) diverges as $t/R \rightarrow \pi/2$, when the big crunch singularity hits the boundary. Indeed, $\beta$ is interpreted as the expectation value of ${\cal O}$ in the dual CFT, and the bulk analysis predicts that to leading order in $1/N$, $\langle {\cal O}\rangle$ diverges in finite time.

For our purposes it suffices to concentrate on the steepest negative direction of the potential. Fluctuations in orthogonal directions in field space acquire a positive mass and are suppressed. For the $SO(5)$-invariant operator we consider, the most unstable direction comes from the $-\Phi_1^4$ term in \eq{Ocubed}. From now on we focus on the dynamics of $\Phi_1$ as it rolls along a fixed direction in $SU(N)$, i.e. $\Phi_1(x)=\phi(x)U$, with $U$ a constant Hermitian matrix satisfying ${\rm Tr} \,U^2=1$, so that $\phi$ is a canonically normalized scalar field. 
The action for this scalar is then given by
\begin{equation}\label{scaction}
S = \int_{ \mathbb{R} \times S _{3}} d ^4x \sqrt{-g} \left( - \frac{1}{2}(\partial \phi) ^2 - \frac{1}{12} \mathrm{R} ^{(2)} \phi ^2 + \frac{1}{4} \lambda \phi ^4 \right)
\end{equation}
where $\lambda = f/N^2$, $ \mathrm{R} ^{(2)} = 6/R^2$ is the Ricci scalar of the 3-sphere, and the coefficient of the $ \phi ^2$ term is fixed by the conformal symmetry. 

Although this is clearly a huge simplification of the field theory, at the classical level it captures the bulk 
behavior in a surprisingly quantitative way. In particular, it admits the following exact homogeneous classical solution
\be \label{clsol}
\phi (t) = \frac{(2/R^2 \lambda)^{1/2}}{\cos (t/R)}.
\ee
which is analogous to the background evolution of the AdS cosmology considered in \cite{Hertog:2004rz}. Indeed, since $\phi^2$ is identified with $\b$ on the bulk side, the time dependence of this solution agrees with the supergravity prediction\footnote{It follows from Fig 14 (b) in \cite{Craps2007} that for sufficiently small values of $f$ the supergravity analysis predicts $\b \sim 1/\lambda$. The field theory solution \eq{clsol} reproduces this scaling.}, including the fact that the field diverges at $t/R \rightarrow \pi/2$. Although conformal invariance is broken quantum mechanically, it has been shown that quantum corrections do not turn around the potential \cite{Craps2007}.

\subsubsection*{Quantum Evolution}
We have seen that evolution ends in finite time in the classical description of a simplified version of the dual field theory, in agreement with the supergravity result. It has been noted, however, that this conclusion changes dramatically if one considers the quantum mechanics of the dual model (\ref{scaction}). That is, one again concentrates on the homogeneous mode $\phi (t)=x(t)$ only, but one now treats this quantum mechanically. 

In quantum mechanics, to ensure unitarity one constructs a self-adjoint extension of the Hamiltonian. This is done by carefully specifying its domain \cite{Reed:1975uy,Carreau90}. One finds the center of a wave packet follows essentially the classical trajectory and still reaches infinity in finite time, but it bounces off infinity and reappears as a left-moving wave packet \cite{Carreau90}. Hence a quantum mechanical treatment of the homogeneous mode of the dual model appears to suggest that there is an immediate transition from a big crunch to a big bang in the bulk.

However, it has been argued \cite{Hertog:2004rz} that the full field theory evolution is likely to be very different.
Indeed, a quantum mechanical analysis obviously neglects the possibility of particle production. Calculations in similar scalar field theories indicate that this significantly affects the evolution of the homogeneous mode. In particular, in many theories where the scalar field rolls down from the top of its effective potential, particles are produced in great numbers {\it while the field is rolling down}. This phenomenon is often called tachyonic preheating \cite{Kofman:2001rb,Felder:2001kt}. It happens essentially because the negative effective mass term in the potential amplifies long wavelength quantum fluctuations.

Tachyonic preheating is so efficient that in some theories most of the initial potential energy density 
is converted into gradient energy well before the field reaches the true minimum. In theories of this kind a prolonged stage of oscillations of the homogeneous component of the scalar field around the true minimum of the potential does not exist in spontaneous symmetry breaking.

One expects that the tachyonic amplification of fluctuations similarly affects the evolution of the dual field theory description of the AdS cosmologies, in which the supergravity initial data correspond to a homogeneous field theory configuration high up the
potential. In this case the scalar potential does not have a global minimum. However, the quantum mechanical self-adjoint extension -- assuming for now it can be implemented in field theory -- essentially acts as a brick wall at infinity. From a cosmological perspective, one is interested in the dynamics at intermediate times, namely whether a wave packet rolling down the potential bounces back one or several times before the system settles down. To answer this one must quantify the effect of the tachyonic instability, and the non-linear growth of fluctuations that comes with it, on the background evolution. This is the subject of the remainder of this paper.

\section{Dual Model}

\subsection{`Brick Wall' Regularization}

\begin{figure}[htbp]
\centering 
\begin{minipage}{8cm}
\centering
\includegraphics[width=8cm]{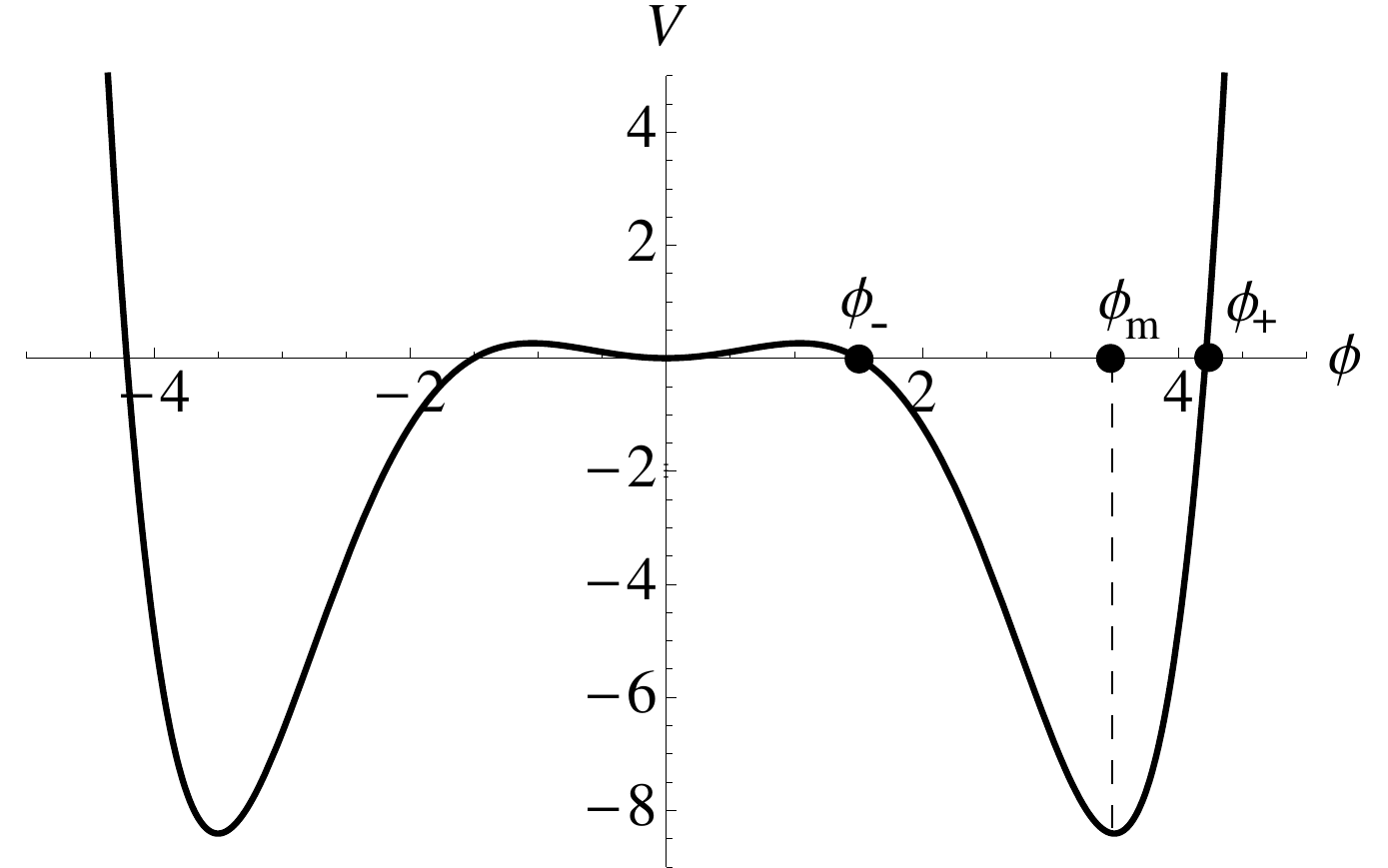}\caption{\label{fig:potential} \small  Regularized scalar potential in the boundary model with $ \lambda = 1$ and $ \gamma = 0.4$. }
\end{minipage}\hspace{0.5cm}
\end{figure}

We have seen that a self-adjoint extension in a quantum mechanical treatment of  (\ref{scaction}) acts as a brick wall at infinity. Perhaps the simplest way to try to implement this at the field theory level is to add a steep regularization term to the potential\footnote{From here onwards we set the AdS radius equal to one, so that the dual field theory lives on $S^3\times R$ where the sphere also has unit radius.},
\begin{equation}\label{eq:potentialReg}
V( \phi) = \frac{1}{2} \phi ^2 - \frac{1}{4} \lambda \phi ^4 + \epsilon \phi ^6
\end{equation}
and to consider the dynamics in the limit $ \epsilon \rightarrow 0 ^{+}$.
It is useful to introduce the parameter $\gamma \equiv 32 \epsilon / 
 \lambda ^{2}$, which can be thought of as the square of a UV regularization length.
In terms of $\gamma$ the potential (\ref{eq:potentialReg}) can be written as
\begin{eqnarray*}
\label{eq:potentialRescaling}
V( \phi)  & = &  \frac{1}{2} \phi ^2 - \frac{1}{4} \lambda \phi ^4 +
\frac{1}{32} \lambda ^2 \gamma \phi ^6  \\
& \equiv &  \lambda ^{-1} V _1(\sqrt{\lambda} \phi)
\end{eqnarray*}
where $V _1$ is defined as $V$ with $\lambda = 1$. For $ \gamma \leq 1$, the potential has two positive roots $ \phi _{ \pm}$ (see Fig \ref{fig:potential}). The global minimum of V on the positive axis is located at $ \phi _{m}$, and we will write $V _{m} \equiv V( \phi _{m})$. 

For $ \gamma \rightarrow 0 ^{+}$, to which we will refer as the \textit{scaling limit}, $ \phi _{-}$ is insensitive to the regulari\-zation and tends to $ \sqrt{2 / \lambda}$. On the contrary, the large $ \phi$ parameters $ \phi _{m}$, $ \phi _{+}$ and $V _{m}$ follow the scaling given by $ \gamma$:
\begin{equation}
\lambda \phi _{+} ^{2} \simeq 8 \gamma ^{-1}, \quad \lambda \phi _{m} ^2 \simeq \frac{16}{3 } \gamma ^{-1} ,  \quad\lambda V _{m} \simeq -\frac{64}{27} \gamma ^{-2}.
\end{equation}

In order to model the background evolution of the AdS cosmologies we consider homogeneous field theory solutions $\bar \phi (t)$ that start at rest at $ \bar{ \phi} = \phi_{-}$. If one neglects fluctuations, the homogeneous field oscillates back and forward between the positive roots of the potential, just as in the quantum mechanical treatment of the unbounded potential discussed above.  The period $T$ of the homogenous field oscillations is  independent of the potential parameters and given by
\begin{equation}\label{eq:period}
T = \sqrt{2} \int _{\phi _{-}} ^{\phi _{+}} \frac{d \phi}{ \sqrt{- V(\phi)}} = \pi
\end{equation}
Hence in the limit $\gamma \rightarrow 0 ^{+}$ the field reaches infinity exactly at $t = \pi/2$, which corresponds to the time at which the big crunch singularity in the bulk reaches the boundary. This suggests that for small $\gamma$ the $\phi ^{6}$ regularization acts very much as a brick wall. Indeed, the interval of time $T_{+}$ during which $ \bar{\phi} \ge  \phi _{m}$ in one oscillation scales as $ \sqrt{\gamma}$,
\be
T _{+} = \sqrt{2} \int _{\phi _{m}} ^{\phi _{+}} \frac{d \phi}{ \sqrt{- V(\phi)}} \stackrel{\gamma \rightarrow 0}{\simeq} \sqrt{\frac{\gamma}{2}}.
\ee

However, the regularization of the boundary potential affects the evolution in the bulk, because it changes the bulk boundary conditions on the timelike boundary. According to the general prescription (\ref{genbc}) these now read
\begin{equation}
\alpha = \lambda \beta - 3 \epsilon \beta ^2.
\end{equation}
For small $\epsilon$ this can only significantly affect the evolution near where the singularity hits the boundary, where $\b$ is large. Nevertheless this suffices to turn the cosmological singularity into a large, stable black hole with scalar hair \cite{Hertog:2005hu}, where the bulk scalar field turned on is dual to the operator (\ref{operator}) in the boundary theory. These hairy black holes, which do not exist for $\a= \lambda \b$ boundary conditions, have been interpreted in the dual theory as thermal excitations about the global negative minimum $V_m$ \cite{Hertog:2005hu}. Hence they are the natural end state of evolution in the bulk corresponding to wave packets rolling down a regularized potential (\ref{eq:potentialReg}).
In the limit $\epsilon \rightarrow 0$ where the global minimum goes to minus infinity the hairy black holes become infinitely large and one recovers the original cosmological solutions  \cite{Hertog:2005hu}. 

Hence in the black hole context there is convincing evidence that the dual system will eventually thermalize. However here we are more interested in the dynamics at intermediate times. Indeed, whether the dual theory predicts a transition from a big crunch to a big bang depends on whether the backreaction of the inhomogeneities prevents an initially homogeneous wave packet rolling down the potential from returning close to its original value after the {\it first} oscillation.

\subsection{Initial Conditions of Perturbations}

Following \cite{Kofman:2001rb, Desroche2005} we set the fluctuations initially in their instantaneous adiabatic vacuum, specified by
\begin{eqnarray}\label{eq:initialSpectrum}
\langle \delta \phi(\vec{k})\, \delta \phi(\vec{k}') \rangle  & \equiv &  \frac{(2 \pi) ^{3}}{2\,\omega _{0} (k)} \delta ^{3}( \vec{k} + \vec{k}')\nonumber \\
\delta \dot{\phi}(\vec{k}) & = & \pm i\, \omega _{0}(k) \delta \phi(\vec{k})
\end{eqnarray}
Because of the unstable $ - \phi^4$ term in the potential, one has $m _{0} ^{2} \equiv V_{,\phi \phi}(\phi _{-}) < 0$. We therefore set
\be \label{omega}
\omega _{0} (k) ^{2} = \left\{ \begin{array}[]{cr}
k ^{2} & k < |m _{0}|\\
k ^{2} + m ^{2}_{0} & k \geq |m _{0}| \end{array} \right.
\ee
to avoid that $\omega _{0}(k)$ is imaginary for $k<|m _{0}|$.
The initialization of the frequency \eq{omega} for the modes with $ k < |m _0|$ is to some extent arbitrary. However, in the limit $\gamma \rightarrow 0$ these modes, which are tachyonic at $t = 0$, constitute only a small part of the ensemble of modes that become tachyonic at some point during the evolution. As part of our numerical study, we have verified that changing the initialization of the $k < |m _0|$ modes does not significantly affect the evolution.

We are initializing random \textit{classical} fluctuations to reproduce the quantum vacuum fluctuations. The consistency of this approach relies on the fact that, as we will show, in the subsequent evolution the occupation numbers become rapidly very large so that the quantum evolution behaves classically, at least during the first stages of the evolution. 

Finally we note that for fixed $\gamma$, changing $\lambda$ is equivalent to rescaling the width of the initial wavepacket. Indeed, defining  $\chi \equiv \sqrt{\lambda} \phi$, the field equation for $ \chi$ is given by
\begin{equation}\label{eq:eomReduced}
\ddot{\chi} - \Delta \chi + \frac{d V _1}{d \chi} = 0
\end{equation}
which is the field equation for $\phi$ with $\lambda = 1$. The homogeneous mode $\bar \chi (t)$ also takes the correct initial value, since $V _1 ( \sqrt{ \lambda} \phi _{-}) = 0$. By contrast  the amplitude of the fluctuations $\delta \chi$ is rescaled by a factor $\sqrt{\lambda}$ with respect to the $\lambda = 1$ model (with $ \hbar = c = 1$):
\begin{equation}
\langle \delta \chi(\vec{k})\, \delta \chi(\vec{k}') \rangle  =  \lambda \frac{(2 \pi) ^{3}}{2\,\omega _{0} (k)} \delta ^{3}( \vec{k} + \vec{k}').
\end{equation}
Hence to study the evolution for a range of values of $\lambda$ one can equivalently change the width of the initial wavepacket, and then go back to the physical variable $ \phi$. 

\subsection{Tachyonic instability}
The effective mass $m ^{2} = V''(\phi)$ of $\phi$ is negative in a range of field values of width $\Delta \phi \sim 1/ \sqrt{ \lambda \gamma}$, extending roughly from $ \phi = \phi _{-} \simeq \sqrt{2/ \lambda}$ to $ \phi \simeq \sqrt{3/ \lambda \gamma}$. The lowest value of $m ^{2}$ is
\begin{equation}\label{eq:tachScale}
m ^{2} _{min} = 1 - \frac{12}{5 \gamma} \stackrel{ \gamma \ll 1}{ \simeq} - \frac{12}{5 \gamma}.
\end{equation}
This value is $ \lambda$-independent, as can be simply deduced from the fact that 
$V_{,\phi \phi} = {V _1}_{,\chi \chi}$.
Modes with momenta $k^2$ between $0$ and $k _{T}^2 \equiv -m ^{2} _{min}$ have an imaginary frequency at some point during the evolution and hence are tachyonic. Decreasing $\gamma$ not only widens the tachyonic band but also amplifies more strongly the modes within the tachyonic range $x \equiv k/k _{T} < 1$. This provides a mechanism for converting potential energy into particles, acting as a friction term for the motion of the homogeneous component of the scalar field.

Eq.(\ref{eq:tachScale}) shows that $\gamma$ acts as a UV regularization length squared. Indeed, the tachyonic instability produces particles with momenta $$ k \lesssim k _{T} \sim \gamma ^{-1/2}$$
irrespective of the value of $\lambda$. However, the details of the spectrum will generally be $ \lambda $ dependent at times sufficiently large for the non-linearities to play a significant role. 

\begin{figure}[h]
\centering 
\begin{minipage}{14cm}
\centering 
\includegraphics[width=8cm]{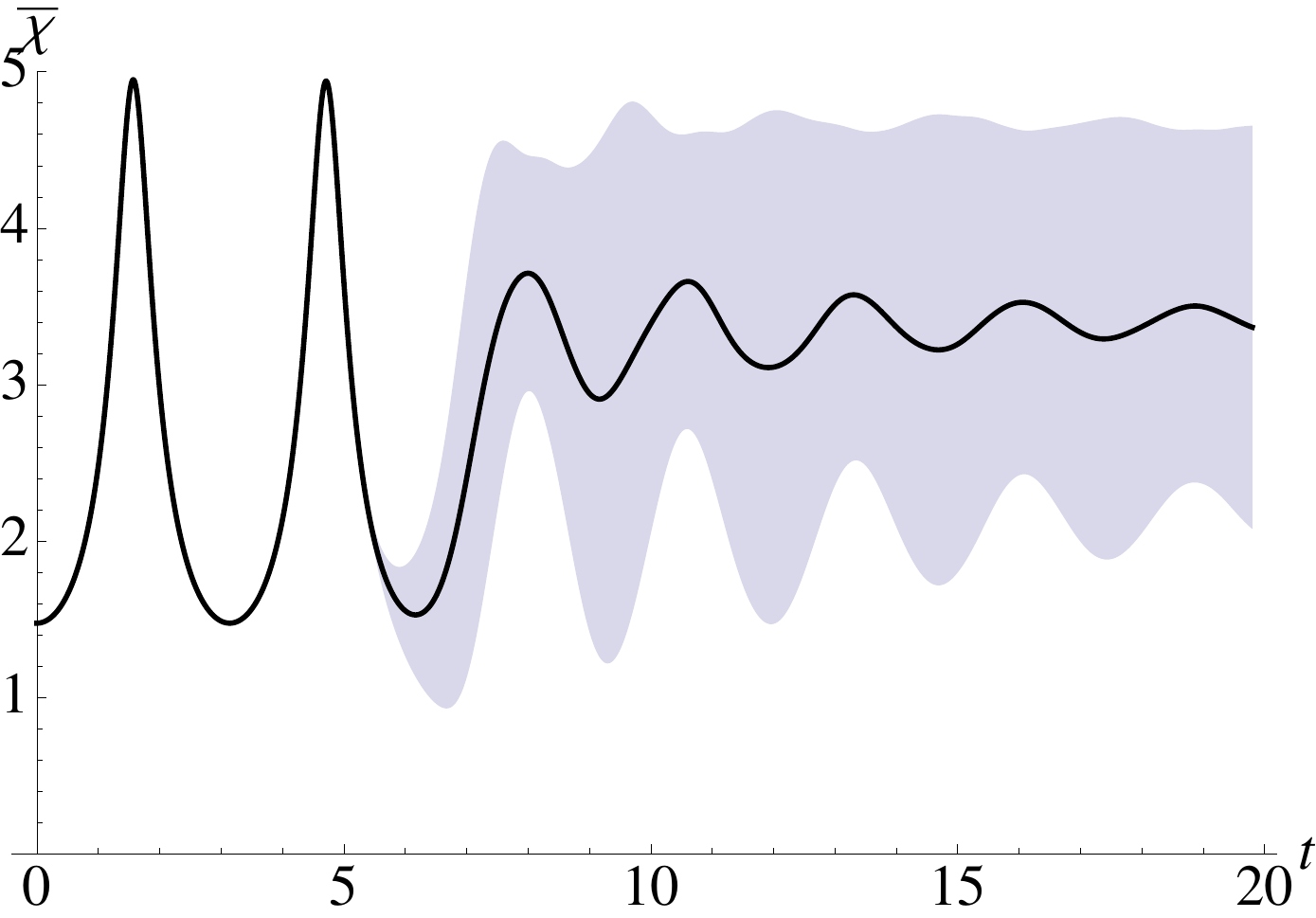}\caption{\label{fig:meanPhi} \small Evolution of the homogeneous component $ \bar{ \chi}(t)$, calculated by averaging the field value over the lattice sites, for $ \lambda/4 =  10 ^{-6}$ and $ \gamma = 0.3$, taking in account the growth of fluctuations $\delta \chi$. The colored band represents the $ \bar{ \chi} \pm \sqrt{ \langle \delta \chi ^2 \rangle}$ zone.}
\end{minipage}
\end{figure}

Systems of this kind have been examined in the context of cosmological preheating and dynamical symmetry breaking \cite{Kofman:2001rb, Desroche2005,Felder:2001kt}. The general conclusion is that the tachyonic amplification is a very efficient mechanism to convert potential energy into particles so that it typically takes only a few oscillations before the homogeneous mode stabilizes around the global minimum. A sample evolution of the homogeneous field in our regularized model (\ref{eq:potentialReg}) of the dual theory is given in Fig \ref{fig:meanPhi}, for $ \lambda/4 =  10 ^{-6}$ and $ \gamma = 0.3$. One sees that with these parameter values the field returns twice to close its original value before backreaction suddenly kicks in and dramatically changes the evolution. Once the homogeneous mode, which is obtained by averaging the field value over the lattice sites, has settled around the global minimum the system thermalizes on a much longer timescale through the gradual transfer of energy into higher harmonic modes. The central question we wish to address here is whether the first oscillation survives in the $\gamma \rightarrow 0 ^{+}$ limit.

To answer this one needs to take in account the evolution of the fluctuations at the non-linear level.
This can be seen as follows. The linearized perturbation equation reads
\begin{equation} \label{linperteq}
\ddot{ \phi _{k}} + (k^2 + m^2(t)) \phi _{k} = 0
\end{equation}
where $m ^2(t) \equiv V _{, \phi \phi}( \bar{ \phi}(t))$. In the absence of the regularization term, one finds that \mbox{$m^2(t) \simeq - 6/ \tau^2$} where $ \tau \equiv \pi/2 - t$ is the time away from the singularity.
Hence for $ \tau \ll 1/k$, \mbox{Eq.(\ref{linperteq})} becomes
\begin{equation}
\ddot{ \phi _{k}} - \frac{6}{ \tau^2}\phi _{k} = 0
\end{equation}

\noindent which is solved by
\begin{equation} \label{eq:linearizedGrowth}
| \phi _{k} ( \tau) | \simeq \frac{| \phi _{k} ( t = 0) | }{ (k \tau )^2}.
\end{equation}
The IR divergence in (\ref{eq:linearizedGrowth}) obviously comes from using the near-singularity behaviour of the classical solution and has no physical meaning. It can be dealt with by imposing an IR cutoff $k _{IR}$ which mimics the actual solution. At a given $ \tau$, one has
\begin{equation}
\langle \delta \phi ^2 \rangle ( \tau) \gtrsim 4 \pi \int _{k _{IR}}^{1/ \tau} dk \, k^2 \frac{ \langle | \phi _{k} (t=0) |^2 \rangle}{ (k \tau)^4} \sim \tau ^{-4} \sim \lambda^2 \bar{ \phi}^4
\end{equation}
where we have restricted the range of integration to the modes that are tachyonic at time $\tau$.
This indicates that while the field rolls down, non-linearities become important when $ \bar{ \phi} \sim 1/ \lambda$. 

\section{Lattice Evolution}

To study the dynamics in the presence of the brick wall regularization, we have developed a numerical code that solves the non-linear Klein--Gordon equation (\ref{eq:eomReduced}) on a 3--dimensional cubical lattice with periodic boundary conditions. We employ the same second order algorithm used in  \cite{Felder:2000hq}. Significant improvements in the numerical performance were made with respect to previous codes. In this section we first describe the numerical setup in more detail and then discuss our results for the classical evolution on the lattice.

\subsection{Numerical Setup}

\subsubsection*{IR cutoff}
The scalar theory (\ref{scaction}) is defined on $ \mathbb{R} \times S ^{3} $. The conformal coupling to the curvature of the \mbox{3--sphere} is crucial to determine the initial conditions for the evolution. On the other hand, in the scaling regime, the tachyonic scale $l _{T}^2 = k _{T} ^{-2} \propto \gamma$ becomes much smaller than $R$.
Hence one expects the IR cutoff given by the radius of the sphere to be unimportant in the $\gamma \rightarrow 0$ limit, so that one can safely work on a toroidal lattice. To verify this is indeed the case we have compared the evolution with the lattice size adjusted so that the momentum gap between the homogeneous mode and the lowest $k$-mode is the same as on the sphere, to the evolution on a very large lattice. It was found that in the scaling regime, the finite volume of the 3--sphere has no effect\footnote{We have also verified that the energy density correlation length (see \cite{Frolov2008}) is much smaller than the IR cutoff.}.

Finally we have verified that, again in the scaling regime, the statistical dependence on the (pseudo--)random initial conditions is negligible if one considers averaged quantities. This provides a consistency check for the semi--classical approach we employ.

\subsubsection*{UV convergence}

\begin{figure}[bhtp]
\centering 
\begin{minipage}{14cm}\centering
\includegraphics[width=8cm]{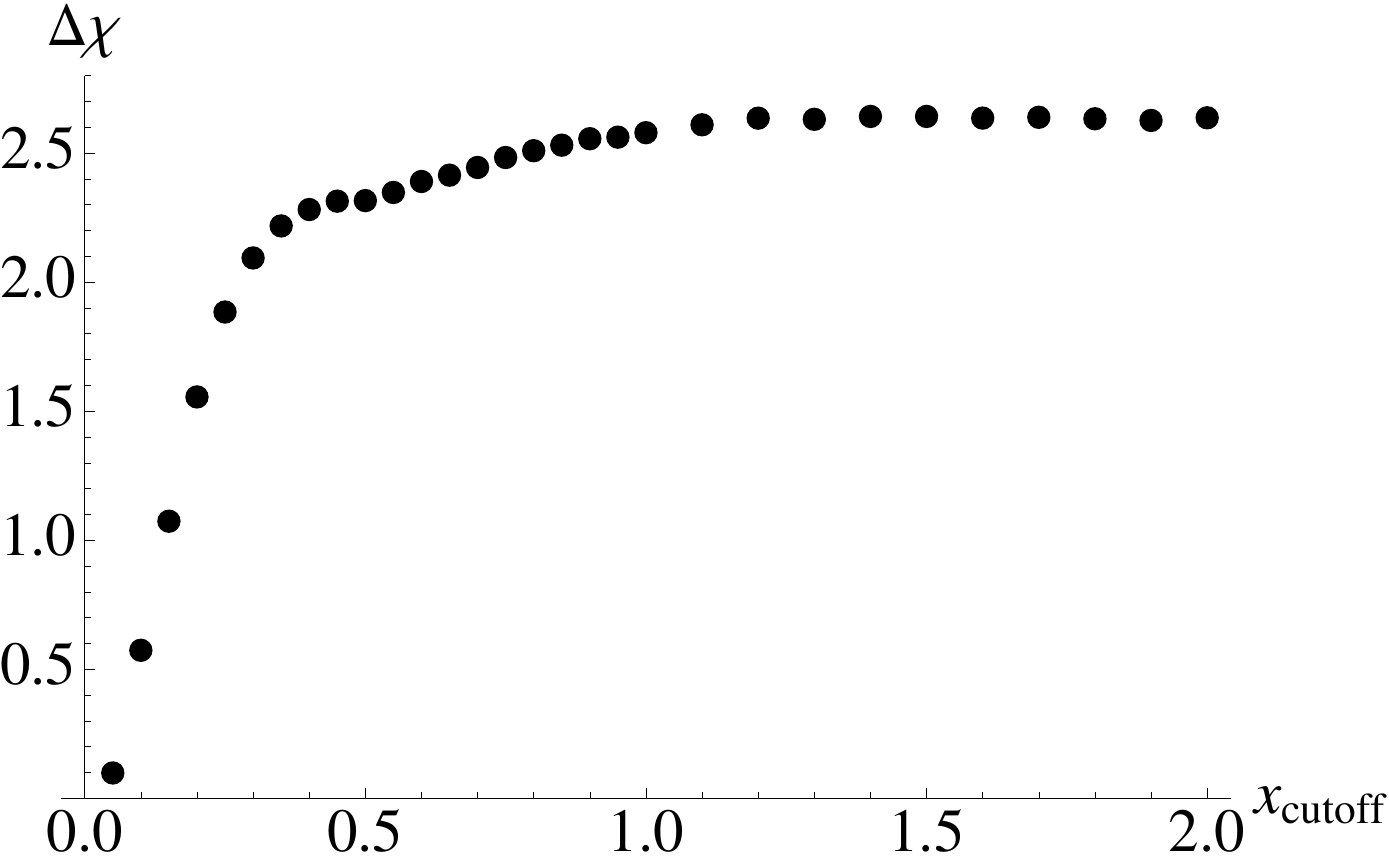}
\caption{\small \label{fig:UVconv1} The `homogeneous' field displacement after the first oscillation, as a function of the lattice UV cutoff, for $ \lambda = 4 \cdot 10^{-3}$, $ \gamma = 10^{-3}$.}
\end{minipage}
\end{figure}

\begin{figure}[htbp]
\begin{minipage}{7.7cm}
\label{eq:deltaPhiDef}
\centering 
\includegraphics[width=7.7cm]{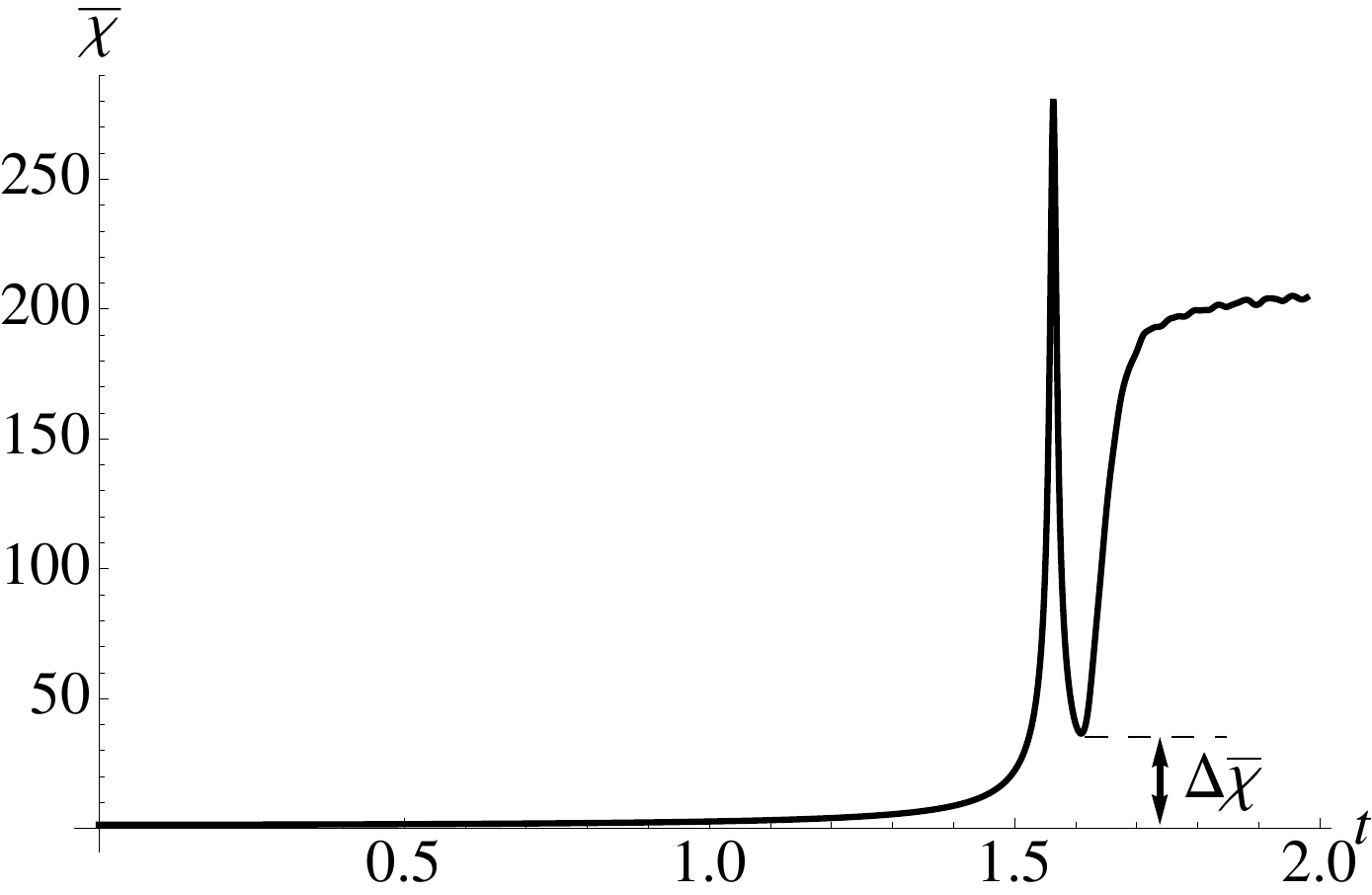}
\end{minipage}\hspace{0.2cm}
\begin{minipage}{7.7cm}
\centering \includegraphics[width=7.7cm]{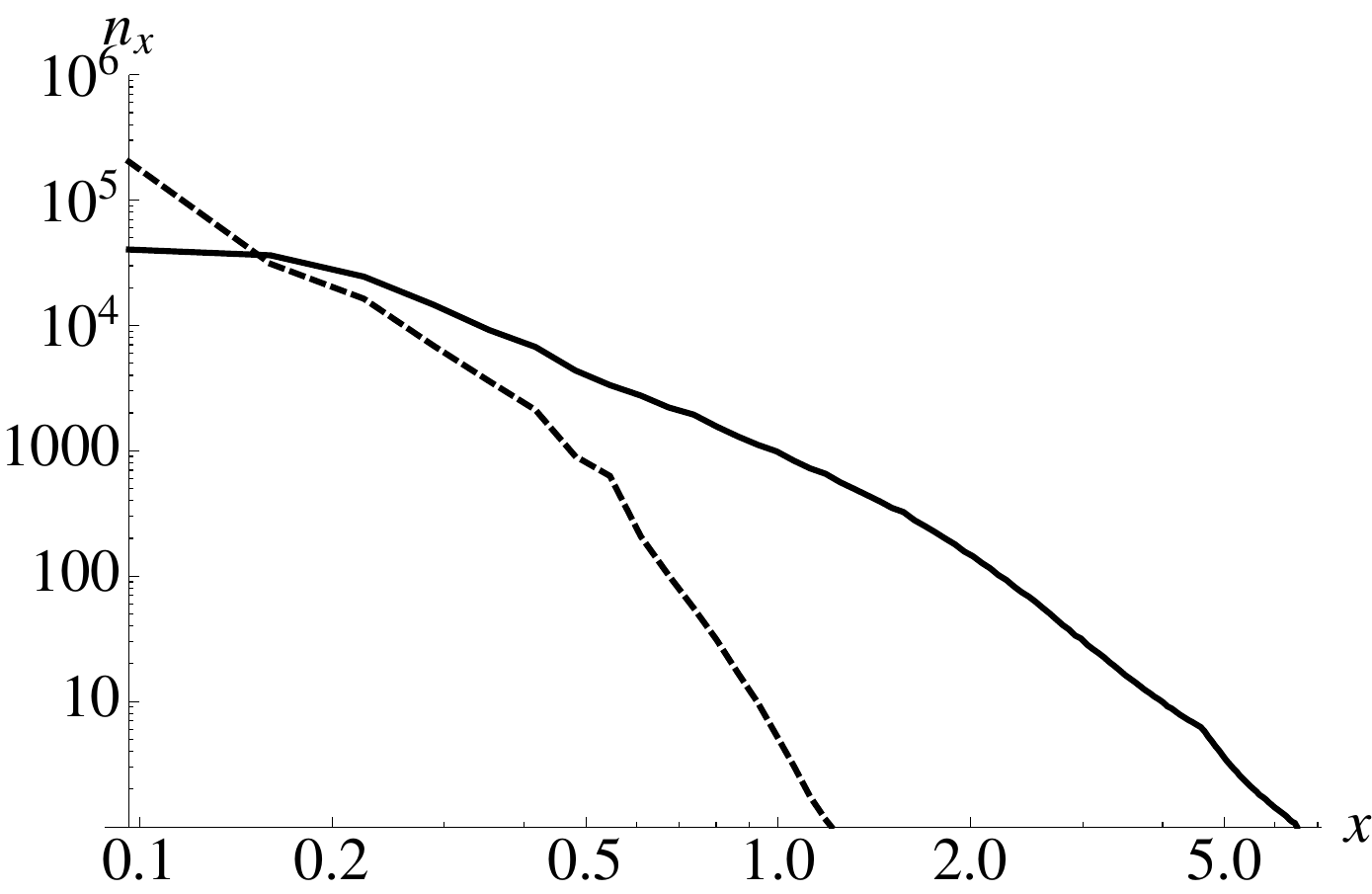}
\end{minipage}
\caption{ \label{fig:deltaphiDef} \small The homogeneous component $ \bar{\chi}$ as a function of time (left), and the mode occupation numbers as a function of $x \equiv k / k _{T}$, at $t \simeq 1.6$ when the first oscillation is completed (dashed) and at $t =2$ (solid). Here $ \lambda /4 = 10 ^{-4}$, $ \gamma = 10 ^{-4}$.}
\end{figure}

We are particularly interested in the early stages of the evolution, such as the dependence of the first oscillation of the field on the parameter values. It turns out that, as expected, in this regime only tachyonic modes with $x \lesssim 1$ are important. This is illustrated in \mbox{Fig \ref{fig:deltaphiDef}} (right panel) where we plot the mode occupation number as a function of $x$ after one oscillation. One sees the occupation numbers fall sharply for 
$x \geq 1$. Hence one expects that a lattice cutoff at $x _{UV} \simeq 2 $ should be sufficient to ensure the UV convergence of the lattice evolution. This turns out to be the case: in Fig \ref{fig:UVconv1} we show that the homogeneous field displacement after one oscillation becomes cutoff independent for $x \gtrsim 2$. 

When the first oscillation ends, the spectrum is essentially cut at $x \simeq 1$. However at later times, when the field relaxes around the minimum, the gradient energy is transferred to higher $x$ modes. This is also shown in Fig \ref{fig:deltaphiDef}, right panel. This is a slow process of which only the first stages are given by classical evolution. However, whereas this is important to understand the ultimate equilibrium state of the field theory, it is not relevant for the `cosmological' question whether there is a bounce in the $\gamma \rightarrow 0$ limit. For this, the field displacement $\Delta \bar \chi$, shown in Fig  \ref{fig:deltaphiDef} (left panel), turns out to be a useful variable.

\subsection{Numerical results}

To evaluate the particle production and in particular its backreaction, we adopt as an indicator the field displacement $ \Delta \bar{ \phi} \equiv \bar{ \phi} ^{(1)} - \phi _{-}$ between the homogeneous field value after the first oscillation, $ \bar{ \phi} ^{(1)}$, and its initial value $ \phi _{-}$ (see Fig \ref{fig:deltaphiDef}). By homogenous field we mean the average of the field value over the lattice sites.
The growth of inhomogeneities can be quantified by evaluating the energy density $ \rho _{g}$ injected in spatial gradients after a single oscillation. These two quantities should naturally be compared to $ (\phi _{m} -\phi _{-})$ and $|V _{m}|$, which determine their upper bounds in the presence of the regularization.

To compare theories with different values of $ \lambda$ it is useful to work with the rescaled quantities $ \Delta \bar{ \chi}  = \sqrt{ \lambda} \,\Delta \bar{ \phi}$ and $ \rho _{ \chi}  = \lambda \rho _{g}$. The advantage of these variables is that the reference quantities $ (\chi _{m} -\chi _{-})$ and $V _{m,1}$ are $ \lambda$--independent. 

Our study covers a wide range of parameters, namely from $ \lambda \simeq 10 ^{-1}$ to $\lambda \simeq 10 ^{-10}$, and from $ \gamma \simeq 10 ^{-1}$ to $ \gamma \simeq 10 ^{-6}$. The lower bounds are imposed by numerical limitations: small values of $\lambda$ require a higher numerical precision to resolve the fluctuations, whereas smaller values of $\gamma$ lead to a more violent and ``tachyonic'' first oscillation that requires a better  discretization.

\subsubsection*{AdS crunch limit: $\gamma$ dependence}

\begin{figure}[thbp]
\begin{minipage}{7.8cm}
\centering 
\includegraphics[width=7.8cm]{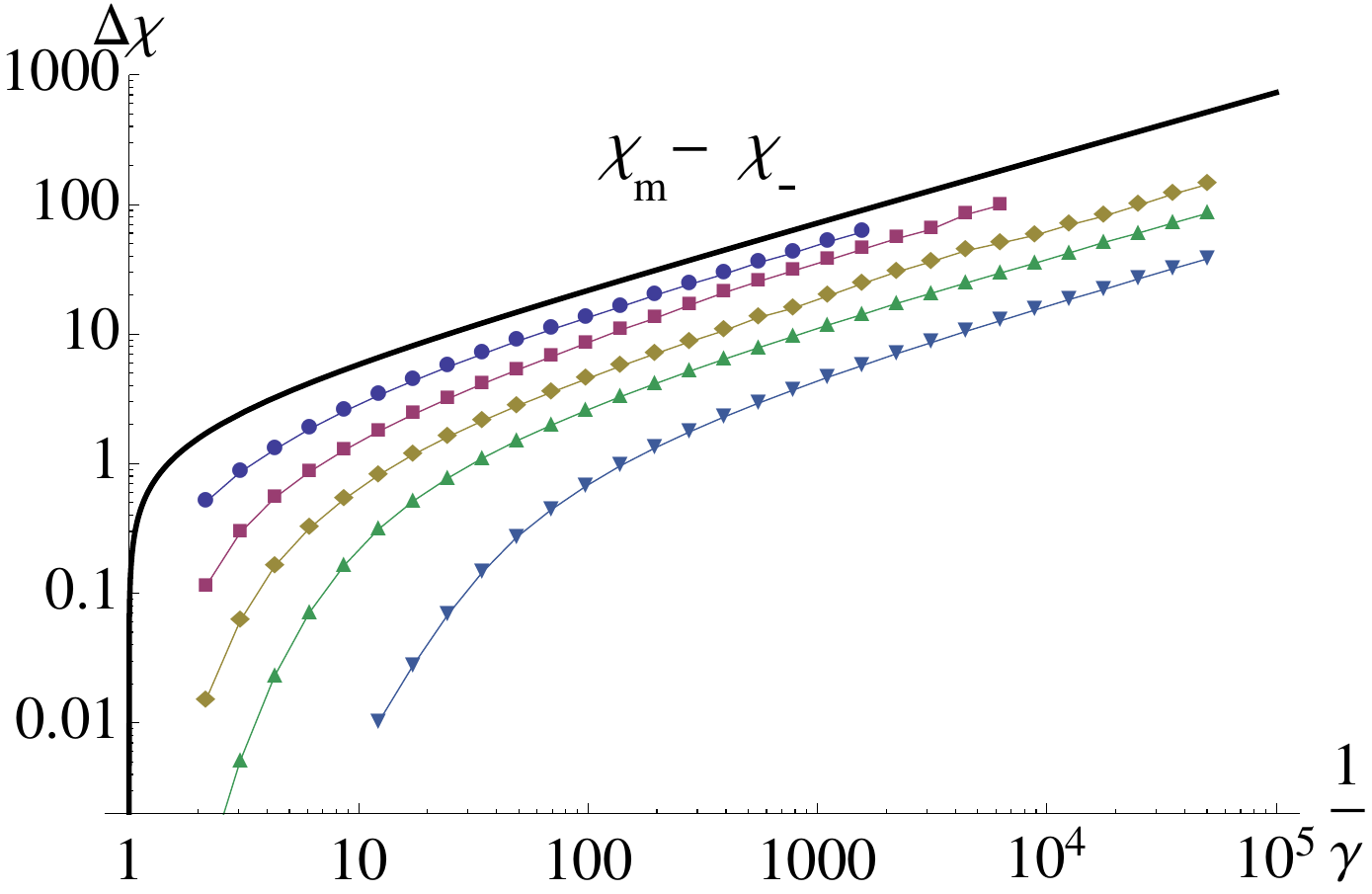}
\end{minipage}\hspace{0.2cm}
\begin{minipage}{7.8cm}
\centering 
\includegraphics[width=7.8cm]{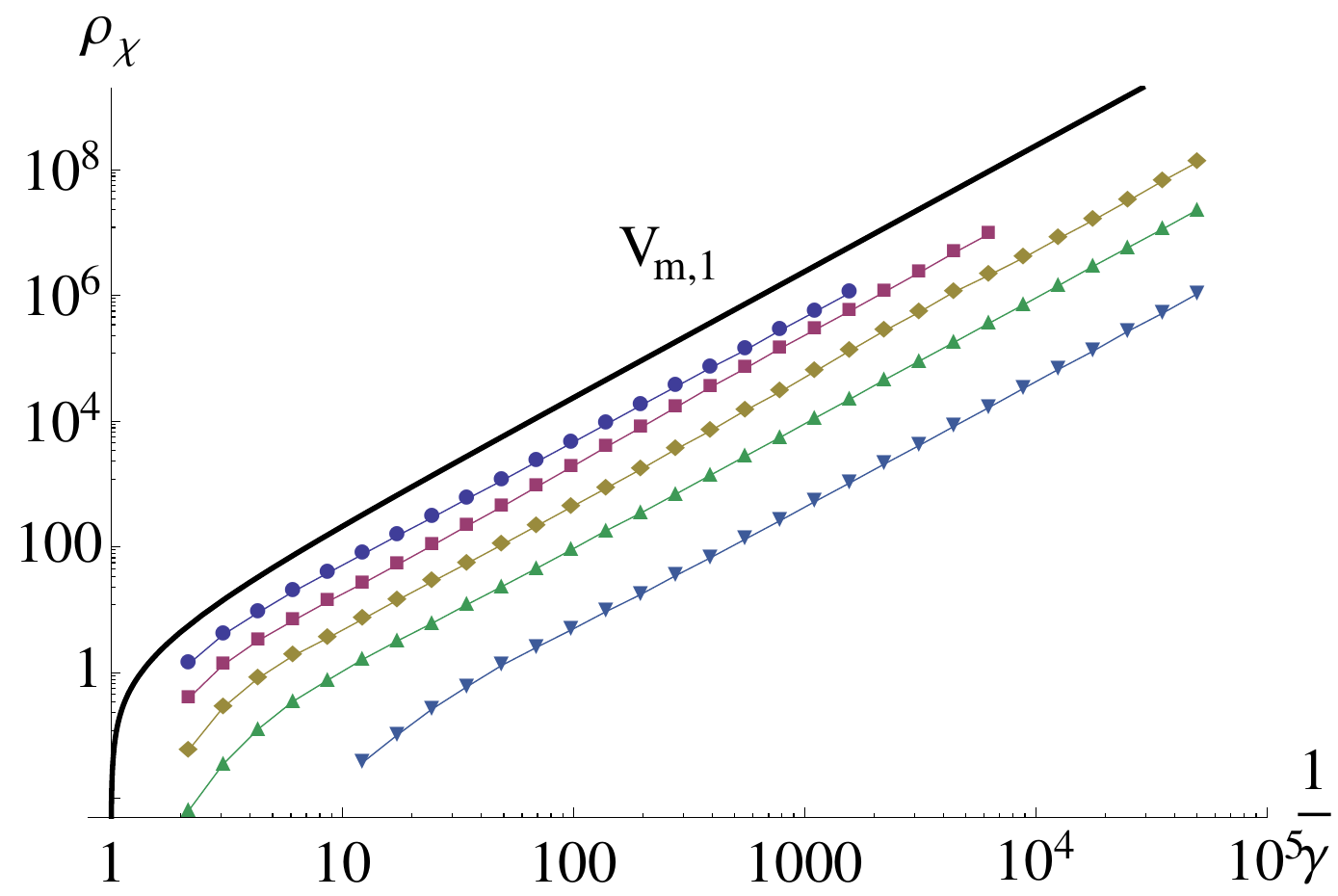}
\end{minipage}
\caption{ \small Homogeneous field displacement (left) and amount of gradient energy density (right) after one oscillation, as  functions of $ \gamma ^{-1}$. The black solid lines represent respectively $ (\chi _{m} -\chi _{-})$ and $V _{m,1}$. From top to bottom, $ \lambda /4  = 10 ^{-1}$, $10 ^{-2}$, $10 ^{-3}$, $10 ^{-4}$, $10 ^{-6}$. }\label{fig:conv}
\end{figure}

For large values of $\gamma$ near $1$, simulations have shown that a few oscillations occur before the homogeneous field reaches its equilibrium value around the true minimum of the potential (see Fig \ref{fig:meanPhi}). At first the fluctuations are in the linear regime, in which their amplitude is essentially multiplied by a constant (which depends on the wavenumber) at each oscillation. However, since the growth of perturbations is exponential, at some point the backreaction becomes important, rapidly dampening the oscillation amplitude.

When one decreases $\gamma$, the potential becomes steeper and deeper. This means that more modes become tachyonic during the first stages of the evolution, and further that the low k modes grow faster. 
We find the energy converted in particles is \textit{not} restored back in the homogeneous modes when the field rolls up again. As a consequence the displacement of the homogeneous field after one oscillation increases when $\gamma$ decreases.

Since for sufficiently small values of $\gamma$ the only relevant scale in the problem is set by $\phi _{m}$, one expects that $\Delta \bar{\phi} \propto \phi _{m}$ and $ \rho _{g} \propto V _{m}$. This is confirmed by the numerical simulations, as shown in Fig \ref{fig:conv}. Hence $ \Delta \bar{ \phi}$ as well as $ \rho _{g}$ follow essentially the scaling that is imposed by the brick wall. The scaling limit is manifest for $ \Delta \bar{\phi} \gtrsim \phi _{-}$. In this regime one has
\begin{equation}\label{eq:gammaScaling}
\Delta \bar{\phi}  \propto \gamma ^{- \frac{1}{2}}, \qquad \rho _{g} \propto \gamma ^{-2}
\end{equation}
The proportionality constants in (\ref{eq:gammaScaling}) depend on $\lambda$ and will be discussed below.

The subsequent evolution depends on the details of the model. For intermediate values of $\gamma$ it takes some time for $ \bar{ \phi}$  to stabilize around $ \phi _{m}$.  This process involves a slow cascading effect whereby the energy is transferred from the tachyonic band to a wider band of modes extending up to wavenumbers of a few times $k _{T}$. 
 
However for small values of $\gamma$ the evolution proceeds in a much more turbulent manner. In this regime the non-linearities are important and even at times  $t \leq T/2$ we find patches with $ \phi \simeq \phi _{m}$ form. The inhomogeneities behave in a local manner till $t \sim T/2$, in line with the findings in \cite{Craps2007}, but for $t \geq T/2$ the evolution proceeds via the expansion and collision of approximately homogeneous bubbles (see also \cite{Barbon2010}). This produces a violent backreaction on $ \bar{ \phi}$ which transfers energy rapidly to high k modes. Consequently, no further large oscillations in $ \bar{ \phi}$ occur. The bubble dynamics for $t \geq T/2$ is heavily dependent on the regularization, and therefore on the kind of boundary conditions we impose at infinity. 

The simulations have also shown that, in the small $\gamma$ regime, a certain amount of tunneling takes place to the true minimum of the potential located at negative $ \phi$. In fact, for small $ \gamma$ the potential barrier is small compared to the scale of the true minimum of the potential, so the potential is essentially flat around $ \phi = 0$. It is therefore not surprising that local energy overdensities let bubbles where the field is negative form. We have checked carefully that this phenomenon is not due to numerical violations of energy conservation. As a matter of fact, this feature of the evolution does not seem crucial for our study. Indeed, an infinite potential barrier at $ \phi = 0$ does not significantly modify the scaling behavior of (\ref{eq:gammaScaling}).

\begin{figure}[t]
\begin{minipage}{7.8cm}
\centering
\includegraphics[width=7.8cm]{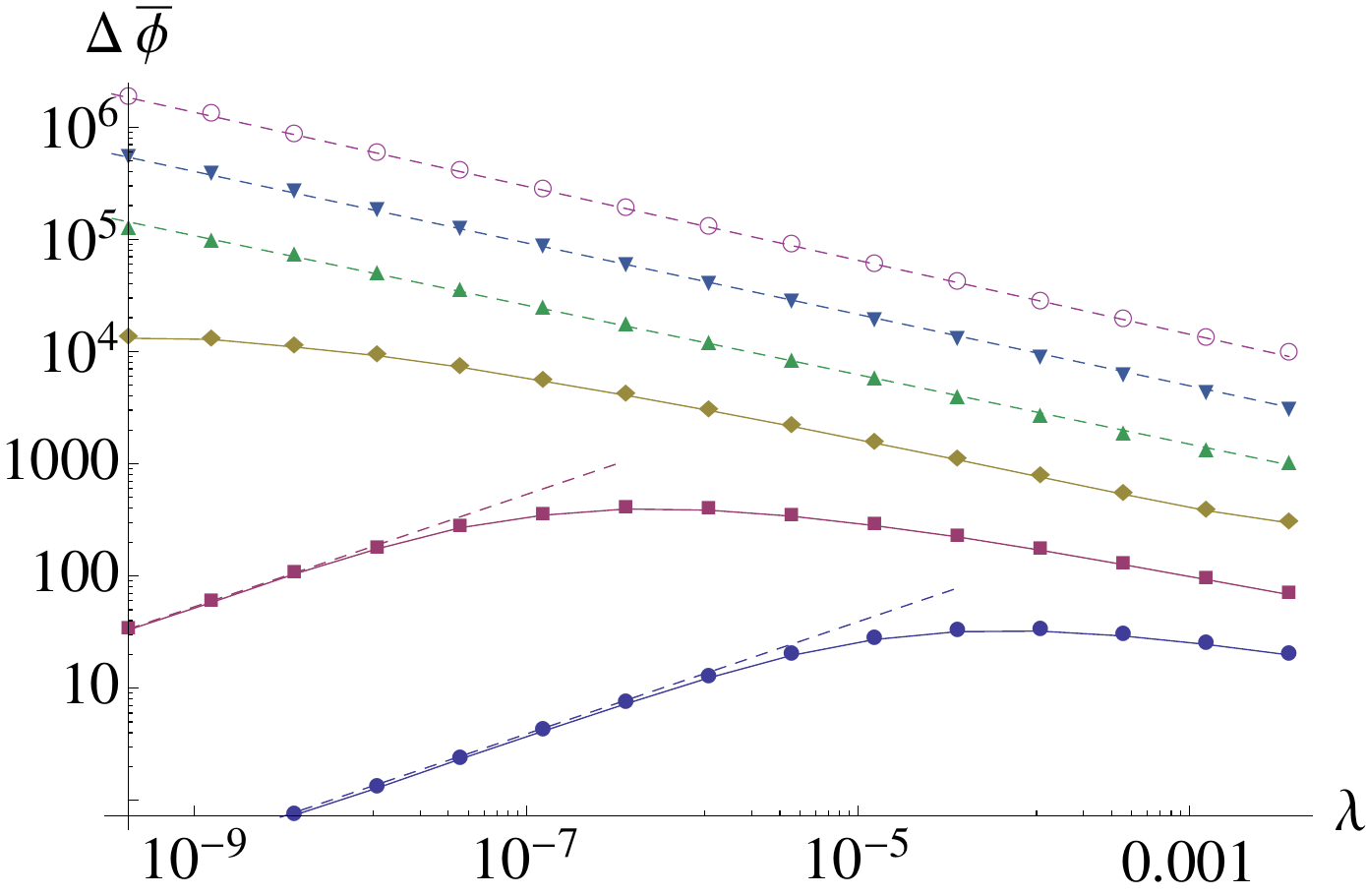}
\end{minipage}\hspace{0.2cm}
\begin{minipage}{7.8cm}
\centering \includegraphics[width=7.8cm]{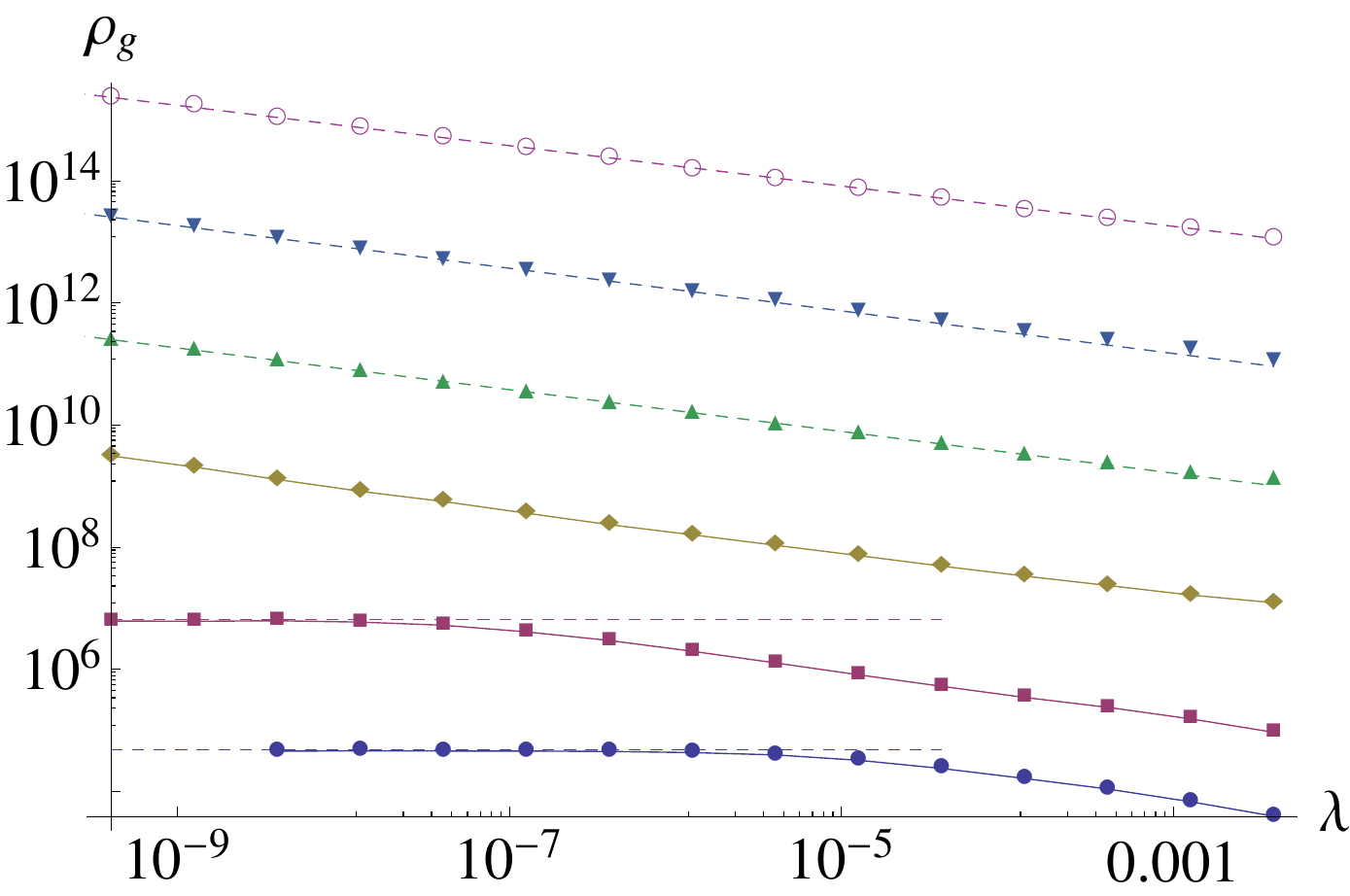}
\end{minipage}
\caption{ \small \label{fig:lambdaDependence}Homogeneous field displacement (left) and amount of gradient energy density  injected (right) after one oscillation, as  functions of $ \lambda$. From top to bottom,  $ \gamma  = 10^{-6},10^{-5},10^{-4},10^{-3},10^{-2},5 \cdot 10^{-2}$. The curves with the two largest values of $\gamma$ exhibit the linear regime for small $\lambda$. The straight dashed lines are, respectively, the $\sqrt{ \lambda}$ and constant fits to the linear analysis discussed in the text. For smaller values of $\gamma$ the backreaction is always large and one obtains the scaling given in Eq (\ref{eq:powerLaws}).}
\end{figure}

\subsubsection*{Non--linearities: $\lambda$ dependence}

For $\lambda \ll \gamma$ and fixed $ \gamma <1$, the fluctuations are small compared to $ \bar{ \phi}$ and behave linearly at least during the first oscillation\footnote{We emphasize that, since the tachyonic instability is an extremely efficient mechanism to produce particles, this regime typically corresponds to values of $ \lambda$ that are very small compared to (an appropriate power of) $ \gamma$. For this reason the regime in which the perturbations behave linearly for the entire duration of the first oscillation has turned out to be numerically inaccessible for $ \gamma \lesssim 10^{-3}$.}. 
At the linear order, both the initial conditions and the evolution of the $ \phi$ inhomogeneities are $ \lambda$ - independent. Consequently we expect no $ \lambda$--dependence for $ \rho _{g} $ in this regime. One should also have
\begin{equation}\label{eq:linear}
\Delta \bar{ \phi} \simeq \frac{ \rho _{g} }{ V _{, \phi}( \phi _{-})} \propto \sqrt{ \lambda}.
\end{equation} 
By contrast, for larger values of $\lambda$ the first oscillation may already involve significant backreaction, with
$ \Delta \bar{\phi} \gg \phi _{-}$. The homogeneous field displacement is then related to $ \rho _{g}$ as
\begin{equation}\label{eq:relation}
\rho _{g} \sim \lambda ( \Delta \bar{ \phi} ) ^4.
\end{equation}
Numerical simulations have confirmed this qualitative behavior. In Fig \ref{fig:lambdaDependence} we show that for $  \lambda \ll \gamma$ the gradient energy $ \rho _{g}$ becomes independent of $\lambda$ and $ \Delta \bar{ \phi}$ scales as $ \sqrt{ \lambda}$, as expected from the linear analysis (\ref{eq:linear}). For larger ratios of $\lambda /\gamma$ both the homogeneous field displacement and the energy density in fluctuations follow a rather precise scaling with $ \lambda$ which appears to be universal for sufficiently small $ \gamma$. The equal spacing between the data sets in Fig \ref{fig:lambdaDependence} corresponds, of course, to the scaling with $\gamma$ given in (\ref{eq:gammaScaling}). 

To summarize, in the scaling limit $ \gamma \rightarrow 0 ^{+}$ and well beyond the linear regime, the quantities of interest behave as
\begin{equation}\label{eq:powerLaws}
\Delta \bar \phi  \propto \frac{ \lambda ^{ \alpha}}{ \gamma ^{ \frac{1}{2}}}, \qquad
\rho _{g}  \propto \frac{ \lambda ^{ \beta}}{ \gamma ^{ 2}}
\end{equation}
with $ \alpha \simeq \beta \simeq -0.33$, which agrees with (\ref{eq:relation}) to a good accuracy.
These exponents are non--trivial and the result of a highly non--linear evolution involving an interplay between tachyonic growth of individual modes and interactions.

The `cosmological' limit corresponds to taking $ \gamma \rightarrow 0 ^{+}$ for \textit{fixed} $ \lambda$. In this limit, it follows from (\ref{eq:powerLaws}) that $\Delta \phi ^{(1)}$ and $\rho _{g}$ both diverge. Consequently, there is no meaningful first oscillation back to a finite value of the field, and therefore no evidence in our dual model for a transition in the bulk from a big crunch to a big bang.

\subsubsection*{Potential renormalization}

A one-loop computation shows that the coupling $ \lambda$ of the unstable $-\phi^4$ potential runs logarithmically. For large field values the Coleman-Weinberg potential is given by
\begin{equation}
V  _{r} \sim - \frac{ \lambda}{ 4 \log{ \frac{ \phi}{ M}}} \phi ^4
\end{equation}
where $M$ is the renormalization scale. The logarithmic running effectively decreases the parametric divergence of the tachyonic scale with $ \gamma$. At  leading order in the logarithmic corrections one finds
\begin{equation}
k _{T} ^2 \propto \frac{ \gamma ^{-1}}{ \log ^2 \gamma}.
\end{equation}
Modeling the self-adjoint extension again by the addition of a $\phi^6$ regularization term we obtain, as a toy model,
\begin{equation} \label{reglog}
V(\phi) = \frac{1}{2} \phi ^2 - \frac{\lambda}{2 \log \left( \frac{ \phi^2 + \phi _0^2}{M ^2} \right) } \phi ^4+ \frac{1}{32} \lambda^2 \gamma \phi^6
\end{equation}
where $ \phi_0$ is added to produce an essentially constant $ \lambda$ for small values of $ \phi$. 

Intuitively, one expects that the running of $\lambda$ leads to logarithmic corrections to the scalings given in (\ref{eq:powerLaws}). A numerical study of (\ref{reglog}) confirms this: the effect of the logarithmic running appears to add only finite $ \gamma$ corrections to the scalings. This means the energy in particles is still determined by the tachyonic scale, and therefore our conclusions remain unchanged. 

\section{Conclusion}

Using the AdS/CFT correspondence, the approach to a big crunch singularity in five-dimensional asymptotically anti-de Sitter spacetimes can be described in a dual field theory by an initially nearly homogeneous wave packet that rolls down an unbounded potential direction of a double trace deformation $-f {\cal O}^2/2$ of ${\cal N}=4$ Super-Yang-Mills (SYM) theory on $ \mathbb{R} \times S ^{3} $, with ${\cal O}$ a trace operator quadratic in the adjoint Higgs scalars. The big crunch singularity in the bulk occurs when the boundary scalar runs to infinity in finite time.

Concentrating on the steepest negative direction of the potential, we have numerically solved for the dual evolution. Consistent quantum evolution requires that one considers a self-adjoint extension of the system. Since this essentially acts as a brick wall at infinity,  we have modeled this by adding a steep regularization term to the unbounded potential. This introduces a UV regulator $\gamma$, and the cosmological limit in the bulk corresponds to taking $\gamma \rightarrow 0^{+}$. 

While the wave packet rolls down the potential, the negative effective mass of the potential amplifies quantum perturbations which therefore behave classically, at least during the first stages of the evolution. 
We find that for small $\gamma$ this converts most of the potential energy of the initial configuration into gradient energy during the first oscillation of the wave packet. In particular, for small $\gamma$ the system tends to a universal `scaling limit' in which the mean field displacement $\Delta \bar{\phi}$ and the gradient energy $\rho _{g}$ after one oscillation scale as (an appropriate power of) the UV regulator itself. This means that, in the cosmological $\gamma \rightarrow 0^{+}$ limit, the boundary scalar does not return close to its original value after the first oscillation. Instead the gradient energy diverges, indicating there is no transition in the bulk from a big crunch to a big bang in the models considered here. In the boundary theory, our simulations show a sudden phase of bubble formation at large field, when some regions of space bounce back while others are still going down. 

A different way to implement self-adjoint boundary conditions in field theory, based on an extension of the domain of the scalars involved, has been explored in \cite{Craps2007} and more recently in \cite{Turok2010}.
There is some evidence \cite{Hertog} this allows for a smoother dynamical evolution across the singularity and in particular that this avoids the breakdown of the local behavior of the perturbations near the singularity. Hence a violent phase of bubble dynamics like what we observed in our simulations may not occur with boundary conditions of this kind. We note, however, that in the models we have considered we find that, in the scaling regime, perturbations are in the non-linear regime already while the wave packet rolls down. It seems plausible that this is largely independent of the nature of the boundary conditions at infinity. One will therefore have to take in account non-linearities in order to determine conclusively whether the boundary scalar can return close to its original value after the first oscillation with different boundary conditions at infinity. 

A possible exception are models where the coupling $\lambda$ governing the potential instability is significantly smaller than the UV regularization scale $\gamma$ for large field values. In this case our simulations indicate that the perturbations remain in the linear regime during the entire first oscillation. The backreaction of particle production can then be limited, at least in a certain range of parameter values. In the models discussed here, we have seen it is not possible to maintain $\lambda < \gamma$ in the cosmological regime involving small $\gamma$. Indeed, although the coupling of the unstable double trace deformation runs logarithmically and is asymptotically free, we find that in order to limit the backreaction $\lambda$ would have to go to zero in the UV faster than this. Whether this can be realized in a quantum treatment of the three-dimensional models discussed in \cite{Craps2009} remains to be seen.

Even though in the regularized models, the expectation value of ${\cal O}$ rapidly stabilizes around the global minimum of the potential,
the spectrum of produced particles at early times is essentially limited to the low tachyonic band of momenta. Subsequently the system thermalizes, due to the self-interaction of the field. However this takes place on much longer time scales and cannot be described by the classical simulations discussed here. The approach to a thermal final state in the boundary theory nevertheless provides an interesting testing ground of the AdS/CFT correspondence in the presence of multitrace deformations. Some aspects of this will be discussed elsewhere \cite{Battarra10b}.

\section*{Acknowledgments}
TH thanks Ben Craps and Neil Turok for stimulating discussions and collaboration on related problems. We also thank Ben Craps for helpful comments on a draft of this paper. This work is supported in part by the ANR (France) under grant ANR-09-BLAN-0157.

\addcontentsline{toc}{chapter}{Bibliography}
  \bibliographystyle{utphys}
  \nocite{Hertog:2004ns,Maldacena:1997re,Elitzur:2005kz,Gasperini1993}
    \bibliography{AdSCrunch7}

\providecommand{\href}[2]{#2}\begingroup\raggedright\begin{thebibliography}{10}

\bibitem{Gasperini1993}
M.~{G}asperini and G.~{V}eneziano, ``{Pre - big bang in string cosmology},''
  \href{http://dx.doi.org/10.1016/0927-6505(93)90017-8}{{\em {A}stropart.
  {P}hys.} {\bf 1} (1993)  317--339},
\href{http://arxiv.org/abs/hep-th/9211021}{{\tt arXiv:hep-th/9211021}}.

\bibitem{Khoury2002}
J.~{K}houry, B.~A. {O}vrut, N.~{S}eiberg, P.~J. {S}teinhardt, and N.~{T}urok,
  ``{From big crunch to big bang},''
  \href{http://dx.doi.org/10.1103/PhysRevD.65.086007}{{\em {P}hys. {R}ev.} {\bf
  D65} (2002)  086007},
\href{http://arxiv.org/abs/hep-th/0108187}{{\tt arXiv:hep-th/0108187}}.

\bibitem{Maldacena:1997re}
J.~M. {M}aldacena, ``{The large N limit of superconformal field theories and
  supergravity},'' {\em {A}dv. {T}heor. {M}ath. {P}hys.} {\bf 2} (1998)
  231--252,
\href{http://arxiv.org/abs/hep-th/9711200}{{\tt arXiv:hep-th/9711200}}.

\bibitem{Hertog:2004rz}
T.~{H}ertog and G.~T. {H}orowitz, ``{Towards a big crunch dual},'' {\em {JHEP}}
  {\bf 07} (2004)  073,
\href{http://arxiv.org/abs/hep-th/0406134}{{\tt arXiv:hep-th/0406134}}.

\bibitem{Hertog:2005hu}
T.~{H}ertog and G.~T. {H}orowitz, ``{Holographic description of AdS
  cosmologies},'' {\em {JHEP}} {\bf 04} (2005)  005,
\href{http://arxiv.org/abs/hep-th/0503071}{{\tt arXiv:hep-th/0503071}}.

\bibitem{Craps2007}
B.~{C}raps, T.~{H}ertog, and N.~{T}urok, ``{Quantum Resolution of Cosmological
  Singularities using AdS/CFT},''
\href{http://arxiv.org/abs/0712.4180}{{\tt arXiv:0712.4180 [hep-th]}}.

\bibitem{Elitzur:2005kz}
S.~{E}litzur, A.~{G}iveon, M.~{P}orrati, and E.~{R}abinovici, ``{Multitrace
  deformations of vector and adjoint theories and their holographic duals},''
  {\em {JHEP}} {\bf 02} (2006)  006,
\href{http://arxiv.org/abs/hep-th/0511061}{{\tt arXiv:hep-th/0511061}}.

\bibitem{Barbon2010}
J.~L.~F. Barbon and E.~Rabinovici, ``{Holography of AdS vacuum bubbles},''
  \href{http://dx.doi.org/10.1007/JHEP04(2010)123}{{\em JHEP} {\bf 04} (2010)
  123},
\href{http://arxiv.org/abs/1003.4966}{{\tt arXiv:1003.4966 [hep-th]}}.

\bibitem{Bernamonti2009}
A.~{B}ernamonti and B.~{C}raps, ``{D-Brane Potentials from Multi-Trace
  Deformations in AdS/CFT},''
  \href{http://dx.doi.org/10.1088/1126-6708/2009/08/112}{{\em {JHEP}} {\bf 08}
  (2009)  112},
\href{http://arxiv.org/abs/0907.0889}{{\tt arXiv:0907.0889 [hep-th]}}.

\bibitem{Asnin2009}
V.~{A}snin, E.~{R}abinovici, and M.~{S}molkin, ``{On rolling, tunneling and
  decaying in some large N vector models},''
  \href{http://dx.doi.org/10.1088/1126-6708/2009/08/001}{{\em {JHEP}} {\bf 08}
  (2009)  001},
\href{http://arxiv.org/abs/0905.3526}{{\tt arXiv:0905.3526 [hep-th]}}.

\bibitem{Hertog:2004ns}
T.~{H}ertog and G.~T. {H}orowitz, ``{Designer gravity and field theory
  effective potentials},''
  \href{http://dx.doi.org/10.1103/PhysRevLett.94.221301}{{\em {P}hys. {R}ev.
  {L}ett.} {\bf 94} (2005)  221301},
\href{http://arxiv.org/abs/hep-th/0412169}{{\tt arXiv:hep-th/0412169}}.

\bibitem{Hertog:2004dr}
T.~{H}ertog and K.~{M}aeda, ``{Black holes with scalar hair and asymptotics in
  N = 8 supergravity},'' {\em {JHEP}} {\bf 07} (2004)  051,
\href{http://arxiv.org/abs/hep-th/0404261}{{\tt arXiv:hep-th/0404261}}.

\bibitem{Henneaux:2004zi}
M.~{H}enneaux, C.~{M}artinez, R.~{T}roncoso, and J.~{Z}anelli,
  ``{Asymptotically anti-de Sitter spacetimes and scalar fields with a
  logarithmic branch},''
  \href{http://dx.doi.org/10.1103/PhysRevD.70.044034}{{\em {P}hys. {R}ev.} {\bf
  D70} (2004)  044034},
\href{http://arxiv.org/abs/hep-th/0404236}{{\tt arXiv:hep-th/0404236}}.

\bibitem{Witten:2001ua}
E.~{W}itten, ``{Multi-trace operators, boundary conditions, and AdS/CFT
  correspondence},''
\href{http://arxiv.org/abs/hep-th/0112258}{{\tt arXiv:hep-th/0112258}}.

\bibitem{Berkooz:2002ug}
M.~{B}erkooz, A.~{S}ever, and A.~{S}homer, ``{Double-trace deformations,
  boundary conditions and spacetime singularities},'' {\em {JHEP}} {\bf 05}
  (2002)  034,
\href{http://arxiv.org/abs/hep-th/0112264}{{\tt arXiv:hep-th/0112264}}.

\bibitem{Reed:1975uy}
M.~{R}eed and B.~{S}imon, {\em {Methods of Modern Mathematical Physics. 2.
  Fourier Analysis, Selfadjointness}}.
\newblock New York 1975, 361p.

\bibitem{Carreau90}
M.~{C}arreau, E.~{F}arhi, S.~{G}utmann, and P.~F. {M}ende, ``{The Functional
  Integral for Quantum Systems with Hamiltonians Unbounded from Below},''
\href{http://dx.doi.org/10.1016/0003-4916(90)90125-8}{{\em {A}nn. {P}hys.} {\bf
  204} (1990)  186--207}.

\bibitem{Kofman:2001rb}
L.~{K}ofman, ``{Tachyonic preheating},''
\href{http://arxiv.org/abs/hep-ph/0107280}{{\tt arXiv:hep-ph/0107280}}.

\bibitem{Felder:2001kt}
G.~N. {F}elder, L.~{K}ofman, and A.~D. {L}inde, ``{Tachyonic instability and
  dynamics of spontaneous symmetry breaking},''
  \href{http://dx.doi.org/10.1103/PhysRevD.64.123517}{{\em {P}hys. {R}ev.} {\bf
  D64} (2001)  123517},
\href{http://arxiv.org/abs/hep-th/0106179}{{\tt arXiv:hep-th/0106179}}.

\bibitem{Desroche2005}
M.~{D}esroche, G.~N. {F}elder, J.~M. {K}ratochvil, and A.~D. {L}inde,
  ``{Preheating in new inflation},''
  \href{http://dx.doi.org/10.1103/PhysRevD.71.103516}{{\em {P}hys. {R}ev.} {\bf
  D71} (2005)  103516},
\href{http://arxiv.org/abs/hep-th/0501080}{{\tt arXiv:hep-th/0501080}}.

\bibitem{Felder:2000hq}
G.~N. {F}elder and I.~{T}kachev, ``{LATTICEEASY: A program for lattice
  simulations of scalar fields in an expanding universe},''
  \href{http://dx.doi.org/10.1016/j.cpc.2008.02.009}{{\em {C}omput. {P}hys.
  {C}ommun.} {\bf 178} (2008)  },
\href{http://arxiv.org/abs/hep-ph/0011159}{{\tt arXiv:hep-ph/0011159}}.

\bibitem{Frolov2008}
A.~V. {F}rolov, ``{DEFROST: A New Code for Simulating Preheating after
  Inflation},'' \href{http://dx.doi.org/10.1088/1475-7516/2008/11/009}{{\em
  {JCAP}} {\bf 0811} (2008)  009},
\href{http://arxiv.org/abs/0809.4904}{{\tt arXiv:0809.4904 [hep-ph]}}.

\bibitem{Turok2010}
N.~{T}urok, ``{Holographic Singularity Resolution},''.
  \url{http://pirsa.org/10060026/}.

\bibitem{Hertog}
B.~{C}raps, T.~{H}ertog, and N.~{T}urok, ``Self-adjoint extensions in field
  theory,'' {\em unpublished manuscript}  .

\bibitem{Craps2009}
B.~{C}raps, T.~{H}ertog, and N.~{T}urok, ``{A multitrace deformation of ABJM
  theory},'' \href{http://dx.doi.org/10.1103/PhysRevD.80.086007}{{\em {P}hys.
  {R}ev.} {\bf D80} (2009)  086007},
\href{http://arxiv.org/abs/0905.0709}{{\tt arXiv:0905.0709 [hep-th]}}.

\bibitem{Battarra10b}
L.~{B}attarra and T.~{H}ertog {\em in preparation}  .

\end{thebibliography}\endgroup

\end{document}